\documentstyle[emulateapj]{article}

\begin{document}

\lefthead{TANIGUCHI, MURAYAMA, \& MOURI}
\righthead{DUSTY TORI OF SEYFERT NUCLEI}

\title{DUSTY TORI OF SEYFERT NUCLEI}
\author{Yoshiaki Taniguchi\altaffilmark{1},
        Takashi Murayama\altaffilmark{1}, and
        Hideaki Mouri\altaffilmark{2}}

\submitted{Proceedings of the Symposium: ``Japan-Germany Workshop on AGN
           and the X-ray Background'' held in Tokyo, Japan, November 1-3,
           1999, eds. T.\ Takahashi and H.\ Inoue, ISAS Report}

\altaffiltext{1}{Astronomical Institute, Graduate School of Science, 
       Tohoku University, Aramaki, Aoba, Sendai 980-8578, Japan} 

\altaffiltext{2}{Meteorological Research Institute,
       1-1 Nagamine, Tsukuba 305-0052, Japan}

\section{INTRODUCTION}
Dusty tori around active galactic nuclei (AGNs) play an important role
in the classification of Seyfert galaxies.
(Antonucci \& Miller 1985; see also Antonucci 1993 for a review).
Seyfert galaxies observed from a face-on view of the torus are
recognized as type 1 Seyferts (S1s) while those observed from an edge-on
view are recognized as type 2 Seyferts (S2s).
Therefore, physical properties of dusty tori are of great interest.
We briefly introduce three statistical studies investigating
properties of dusty tori;
1) physical sizes of dusty tori based on water-vapor maser emission
   (Taniguchi \& Murayama 1998),
2) ionization condition of the inner wall of tori based on
   high-ionization emission lines
   (Murayama \& Taniguchi 1998a,b),
and
3) viewing angle toward dusty tori based on mid-infrared color
   (Murayama, Mouri, \& Taniguchi 2000).
Please see references for detailed discussion.

\section{Dusty Tori of Seyfert Nuclei Posed by
         the Water Vapor Maser Emission}

\subsection{Water Vapor Maser Emission in Active Galactic Nuclei}
The recent VLBI/VLBA measurements of the H$_2$O maser emission
of the nearby AGNs, NGC 1068
(Gallimore et al. 1996; Greenhill et al. 1996; Greenhill \& Gwinn 1997), 
NGC 4258 (Miyoshi et al. 1995; Greenhill et al. 1995a, 1995b), 
and NGC 4945 (Greenhill, Moran, \& Herrnstein 1997),
have shown that the masing clouds
are located at distances of $\sim$ 0.1 -- 1 pc from the nuclei.
These distances are almost comparable to those of molecular/dusty tori
which are the most important ingredient to explain the observed
diversity of AGN (Antonucci \& Miller 1985; Antonucci 1993).
It is therefore suggested that the masing clouds reside
in the tori themselves (e.g., Greenhill et al. 1996).
Therefore, the H$_2$O maser emission provides a useful tool to study 
physical properties of dusty tori which are presumed to be the fueling agent
onto the supermassive black hole (cf. Krolik \& Begelman 1988; 
Murayama \& Taniguchi 1997).

\subsection{A Statistical Size of the Dusty Tori Inferred
           from the Frequency of Occurrence of H$_2$O Masers}

The recent comprehensive survey of the H$_2$O maser emission for 
$\sim$ 350 AGNs by Braatz et al. (1997; hereafter BWH97) has shown 
that the H$_2$O maser emission has not yet been observed in S1s and
that the S2s with the H$_2$O maser emission have the higher \ion{H}{1}
column densities toward the central engine.
It is hence suggested strongly that the maser emission
can be detected only when the dusty torus is viewed from almost
edge-on views.
This is advocated by the ubiquitous presence of so-called the
main maser component whose velocity is close to the systemic
one whenever the maser emission is observed because this component arises from
dense molecular gas clouds along the line of sight between
the background amplifier (the central engine) and us
(see, e.g., Miyoshi et al. 1995; Greenhill et al. 1995b).

Since the high \ion{H}{1} column density is achieved only
when we see the torus within the aspect angle,
$\phi =\tan^{-1} (h/2b)$ (see Figure 1),
we are able to estimate $b$ because the detection rate of
H$_2$O maser, $P_{\rm maser}$, emission can be related to
the aspect angle as,
$P_{\rm maser} = N_{\rm maser}/(N_{\rm maser} + N_{\rm non-maser}) 
= \cos (90\arcdeg - \phi)$ where 
$N_{\rm maser}$ and $N_{\rm non-maser}$ are the numbers of AGN with
the H$_2$O maser emission and without the H$_2$O maser emission, respectively.
This relation gives the outer radius,
$b = h~ [2 \tan (90\arcdeg - \cos^{-1} P_{\rm maser})]^{-1}$.
Table 1 shows that a typical detection rate is $P_{\rm maser} \simeq$ 0.05.
However, this value should be regarded as a lower limit because
some special properties of may be necessary to cause the maser emission
(Wilson 1998). If we take account of new detections of H$_2$O maser emission
from NGC 5793 (Hagiwara et al. 1997) and NGC 3735 (Greenhill et al. 1997b)
which were discovered by two other maser surveys independent from BWH97,
the detection rate may be as high as $\simeq$ 0.1 (Wilson 1998).
Therefore, we estimate $b$ values for the two cases; 1) $P_{\rm maser}$ = 0.05,
and $P_{\rm maser}$ = 0.1.  These two rates correspond to the aspect angles,
$\phi \simeq 2\fdg{}9$ and $\phi \simeq 5\fdg{}7$, respectively.
In Table 2, we give the estimates of $b$ for three cases, 
$a$ = 0.1, 0.5, and 1 pc.
If $a >$ 1 pc, the \ion{H}{1} column density becomes lower than
$10^{23}$ cm$^{-1}$ given $M_{\rm gas} = 10^5 M_\odot$. 
Therefore, it is suggested that the inner radius may be in a range 
between 0.1 pc and 0.5 pc for typical Seyfert nuclei.
The inner radii of the H$_2$O masing regions in NGC 1068, NGC 4258, and NGC
4945 are indeed in this range (Greenhill et al. 1996; Miyoshi et al. 1997;
Greenhill et al. 1997a).
We thus obtain possible sizes of the dusty tori;
($a, b, h$) = (0.1 -- 0.5 pc, 1.67 -- 8.35 pc, 0.33 -- 1.67 pc) 
for $\phi \simeq 5\fdg{}7$, and 
($a, b, h$) = (0.1 -- 0.5 pc, 3.29 -- 16.5 pc, 0.33 -- 1.67 pc) 
for $\phi \simeq 2\fdg{}9$.
All the cases can achieve $N_{\rm HI} > 10^{23}$ cm$^{-1}$,
being consistent with the observations (BWH97). 

\begin{figure*}
\figurenum{1}
\epsscale{1.4}
\plotone{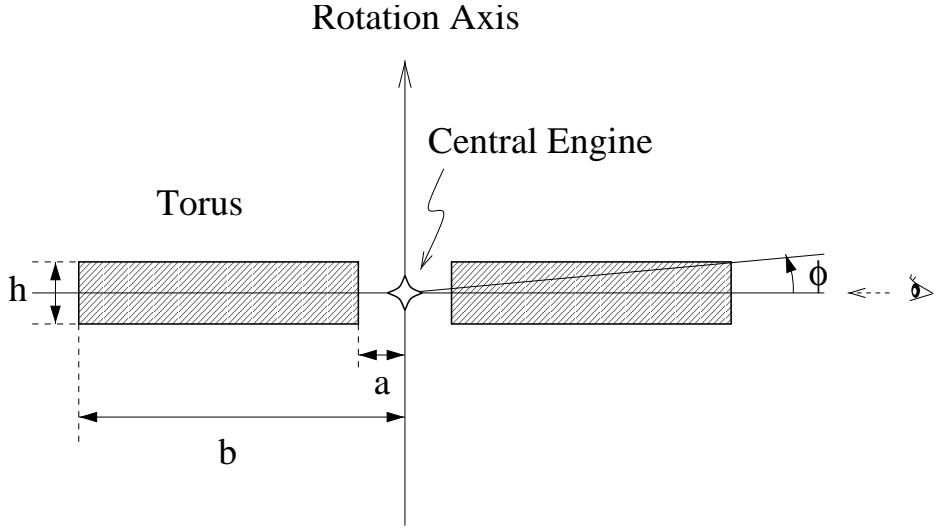}
\caption{The geometry of the dusty torus.
 The torus is a cylinder of dust with a uniform density,
 characterized by the inner radius ($a$), the outer radius ($b$),
 and the full height ($h$). A half-opening angle of the
 torus is thus given as $\phi =\tan^{-1} (h/2b)$. 
}
\end{figure*}

\begin{deluxetable}{cccccc}
\tablenum{1}
\tablecaption{A summary of the detection rates of
        the H$_2$O maser in active galactic nuclei
        studied by Braatz, Wilson, \& Henkel (1997) for the various samples}
\tablehead{
\colhead{Sample} &
\colhead{$N_{\rm maser}$} &
\colhead{$N_{\rm total}$} &
\colhead{$P_{\rm maser}$ (\%)} \nl
}
\startdata
Distance-limited & & & \nl
\hline
All (S1+S2+L) & 15 & 278 & 5.4 \nl
Seyfert (S1+S2) & 10 & 198 & 5.1 \nl
S2 & 10 & 141 & 7.1 \nl
\hline
Magnitude-limited & & & \nl
\hline
All (S1+S2+L) & 13 & 241 & 5.4 \nl
Seyfert (S1+S2) & 8 & 166 & 4.8 \nl
S2 & 8 & 112 & 7.1 \nl
\enddata
\end{deluxetable}

\begin{deluxetable}{ccccccc}
\tablenum{2}
\tablecaption{Geometrical properties of the dusty tori inferred
  from the statistics of the H$_2$O maser emission}
\tablehead{
 & & & \multicolumn{2}{c}{$P_{\rm maser}=0.05$} &
 \multicolumn{2}{c}{$P_{\rm maser}=0.1$} \nl
 & & & \multicolumn{2}{c}{$\phi=2\fdg{}9$} &
 \multicolumn{2}{c}{$\phi=5\fdg{}7$} \nl
 $a$ (pc) & $h$ (pc) &  $r_{\rm hot}$\tablenotemark{a} (pc) &
 $b$ (pc) & $N_{\rm HI}$ (cm$^{-2}$) & $b$ (pc) & $N_{\rm HI}$ (cm$^{-2}$)
}
\startdata
0.1 & 0.33 & 0.43 & 3.29 & $3.3 \times 10^{24}$ &1.67 & $6.5\times 10^{24}$ \nl
0.5 & 1.67 & 2.17 & 16.5 & $1.3\times 10^{23}$ & 8.35 & $2.6\times 10^{23}$ \nl
  1 & 3.30 & 4.30 & 32.9 & $3.3\times 10^{22}$ & 16.7 & $6.5\times 10^{22}$ \nl
\enddata
\tablenotetext{a}{The radius of the hot part in the torus;
                $r_{\rm hot} = a + h$.}
\end{deluxetable}

\section{High-Ionization Nuclear Emission-Line Regions
         on the Inner Surface of Dusty Tori}

\subsection{High-Ionization Emission Lines in Seyfert Galaxies}
Optical spectra of active galactic nuclei (AGN) show often
very high ionization emission lines such as [\ion{Fe}{7}], [\ion{Fe}{10}],
and [\ion{Fe}{14}] (the so-called coronal lines).
According to the current unified models
(Antonucci \& Miller 1985; Antonucci 1993), 
it is generally believed that
a dusty torus surrounds both the central engine and the BLR.
Since the inner wall of the torus is exposed to intense radiation
from the central engine, it is naturally expected that the wall 
can be one of the important sites for the HINER (Pier \& Voit 1995).
If the inner wall is an important site of HINER, 
it should be  expected that the S1s would tend to have 
more intense HINER emission because the inner wall would be 
obscured by the torus itself in S2s.

In order to examine whether or not the S1s tend to have the excess HINER
emission, we study the frequency distributions of the
[\ion{Fe}{7}] $\lambda$6087/[\ion{O}{3}] $\lambda$5007 intensity
ratio between S1s and S2s.
The data were compiled from the literature (Osterbrock 1977, 1985; Koski 1978;
Osterbrock \& Pogge 1985; Shuder \& Osterbrock 1981)
and our own optical spectroscopic  data
of one S1 (NGC 4051) and four S2s (NGC 591, NGC 5695, NGC 5929,
and NGC 5033).
In total, our sample contains 18 S1s and 17 S2s.
The result is shown in Figure 2.
It is shown that the S1s are strong [\ion{Fe}{7}] emitters than the S2s.
In order to verify that this difference is really due to the excess
[\ion{Fe}{7}]
emission, we compare the [\ion{O}{3}] luminosity between the S1s and S2s and
find that the [\ion{O}{3}] luminosity distribution is nearly the same between 
the S1s and the S2s (Figure 3).
Therefore, we conclude that the higher [\ion{Fe}{7}]/[\ion{O}{3}]
intensity ratio in the S1s is indeed due to the
excess [\ion{Fe}{7}] emission  rather than the weaker
[\ion{O}{3}] emission in the S1s.  
The presence of an excess [\ion{Fe}{7}] emission in
S1s can only be explained if
there is a fraction of the inner HINER that cannot be seen in the S2s.
The height of the inner wall is of order 1 pc (Gallimore et al. 1997;
Pier \& Krolik 1992, 1993).
Therefore, given that the torus obscures this HINER from our line of sight,
the effective height of the torus should be significantly higher than 1 pc.

\begin{figure*}
\epsscale{1.0}
\figurenum{2}
\plotone{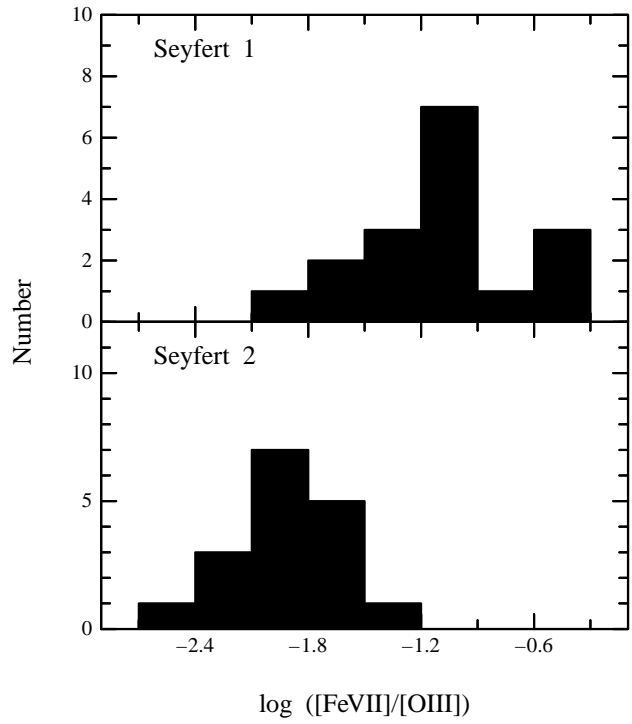}
\caption{Frequency distributions of the
[\ion{Fe}{7}]$\lambda$6087/[\ion{O}{3}]$\lambda$5007
intensity ratio between the S1s and the S2s.
}
\end{figure*}

\begin{figure*}
\figurenum{3}
\plotone{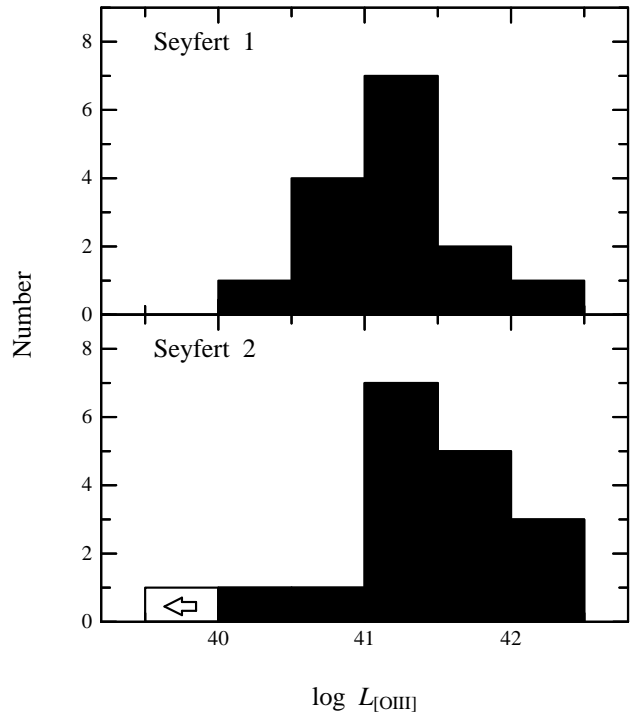}
\caption{Frequency distributions of the [\ion{O}{3}]
$\lambda$5007 luminosity
between the S1s and the S2s.
$H_0=75$ km Mpc$^{-1}$ is assumed.
}
\end{figure*}

\subsection{Three-Component HINER}
Although our new finding suggests strongly that part of
the HINER emission arises from the inner walls of dusty tori,
it is remembered that a number of S2s have the HINER.
In fact, the fraction of Seyfert nuclei with the HINER
is nearly the same between the S1s and the S2s (Osterbrock 1977; Koski 1978).
If the HINER was mostly concentrated in the inner 1 pc region,
we would observe the HINER only in the S1s. 
Therefore the presence of HINER in the S2s implies that there is
another HINER component which has no viewing-angle dependence. 
A typical dimension of such a component is of order 100 pc like
that of the NLR. In addition, it is also known that some Seyfert
nuclei have an extended HINER whose size amounts up to $\sim$ 1
kpc (Golev et al. 1994; Murayama, Taniguchi, \& Iwasawa 1998).
The presence of such extended HINERs is usually explained
as the result of very low-density conditions in the interstellar medium 
($n_{\rm H} \sim 1$ cm$^{-3}$)
makes it possible to achieve higher ionization conditions
(Korista \& Ferland 1989).

The arguments described here suggest strongly that
there are three kinds of HINER; 1) the torus HINER ($r < 1$ pc),
2) the HINER associated with the NLR ($10 < r < 100$ pc), and 
3) the very extended HINER ($r \sim$ 1 kpc).
A schematic illustration of the HINER is shown in Figure 4.
\begin{figure*}
\epsscale{1.2}
\figurenum{4}
\plotone{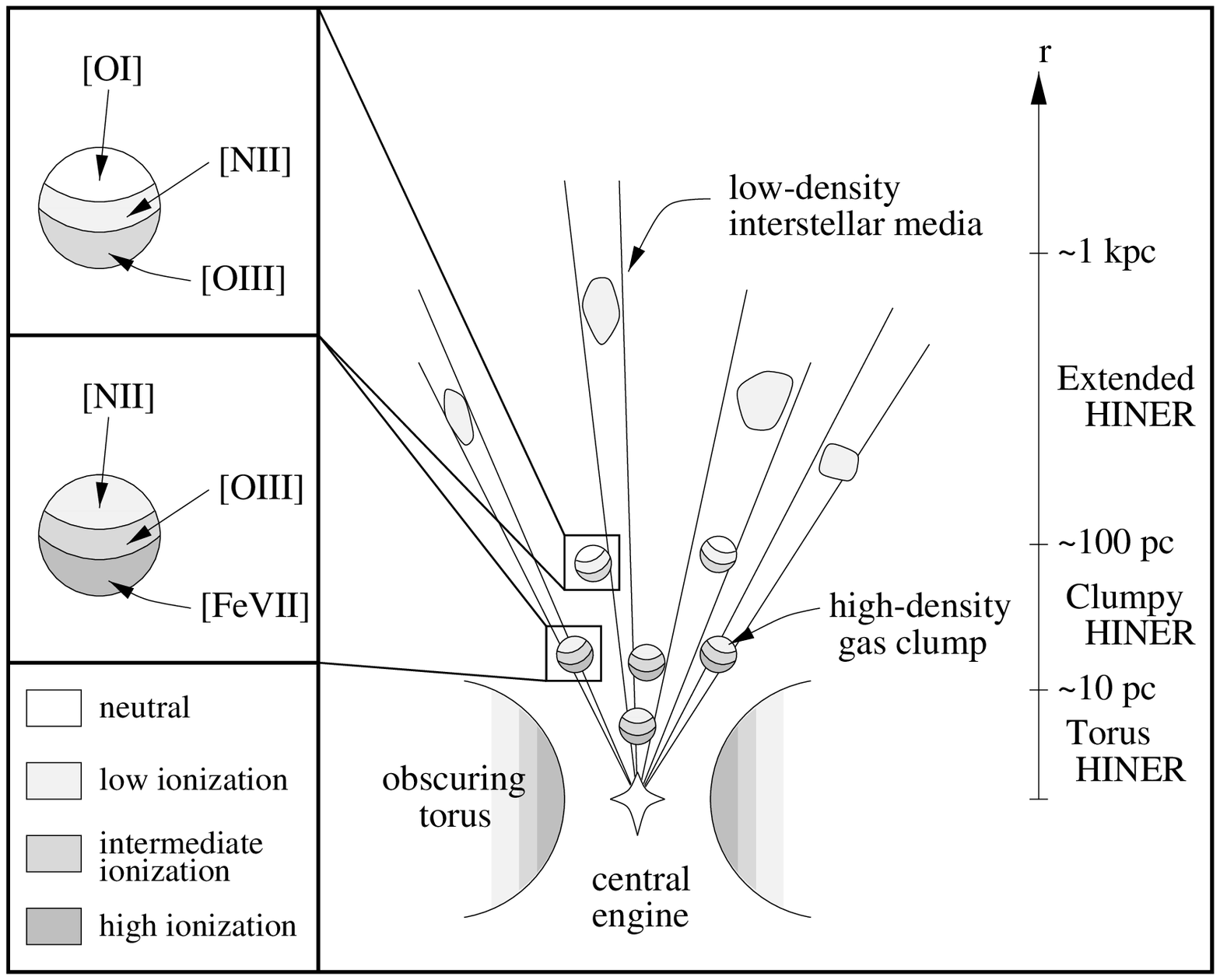}
\caption{A schematic illustration of the three-component model for the HIENR.
Note that the torus HINER consists of many small ionized gas clumps
like the clumpy HINER in the NLR.
}
\end{figure*}

\subsection{Dual-Component Photoionization Calculations for HINER}
Any single-component photoionization models underpredict
higher ionization emission lines (see Murayama \& Taniguchi 1998b
and references therein).
We therefore proceed to construct dual-component models
in which the inner surface of a torus is introduced as
a new ionized-gas  component in addition to the traditional
NLR component with the photoionization code CLOUDY (Ferland 1996).
The single-cloud model suggests that the ionization parameter lies in
the range of $\log U \simeq -1.5$ -- $-2$.
As for the electron density, it is often considered that the inner
edges of tori have higher electron densities, e.g.,
$n_{\rm e} \sim 10^{7\mbox{--}8}$ cm$^{-3}$ (Pier \& Voit 1995).
Because the largest [\ion{Fe}{7}]/[\ion{O}{3}]
ratio of the observed data is $\sim 0.5$, [\ion{Fe}{7}]/[\ion{O}{3}]
of the torus component must be greater than 0.5.
However, we find that ionization-bounded models 
cannot explain the observed large [\ion{Fe}{7}]/[\ion{O}{3}] values
by simply increasing electron densities up to $10^{9}$ cm$^{-3}$.
Further, such very high-density models yield unusually strong [\ion{O}{1}]
emission with respect to [\ion{O}{3}].
We therefore assume ``truncated'' clouds with both large
[\ion{Fe}{7}]/[\ion{O}{3}]
ratios and little low-ionization lines for the HINER torus.
The calculations were stopped at a hydrogen column density
when [\ion{Fe}{7}]/[\ion{O}{3}] $=1$.
We performed photoionization calculations described above and
we finally adopted the model with
$n_{\rm H} = 10^{7.5}$ cm$^{-3}$ and $\log U = -2.0$
representative model for the HINER torus 
with taking [\ion{Fe}{10}]/[\ion{Fe}{7}] ratios predicted by the
calculations into account.

Now we can construct dual-component models combining this torus
component model with the NLR models.
In Figure 5, we present the results of the
dual-component models. Here the lowest dashed line shows the results of the
NLR component models with $\alpha=-1$, $\log U=-2$, as a
function of $n_{\rm H}$ from
1 cm$^{-3}$ to $10^{6}$ cm$^{-3}$. If we allow  the contribution
from the torus component to reach up to $\sim 50$ \% in the Seyferts
with very high  [\ion{Fe}{7}]/[\ion{O}{3}] ratios,
we can explain all the data points without invoking the unusual iron
overabundance.
Note that the majority of objects can be explained by simply
introducing a $\sim 10$ \%
contribution from the HINER torus.
\begin{figure*}
\figurenum{5}
\epsscale{1.2}
\plotone{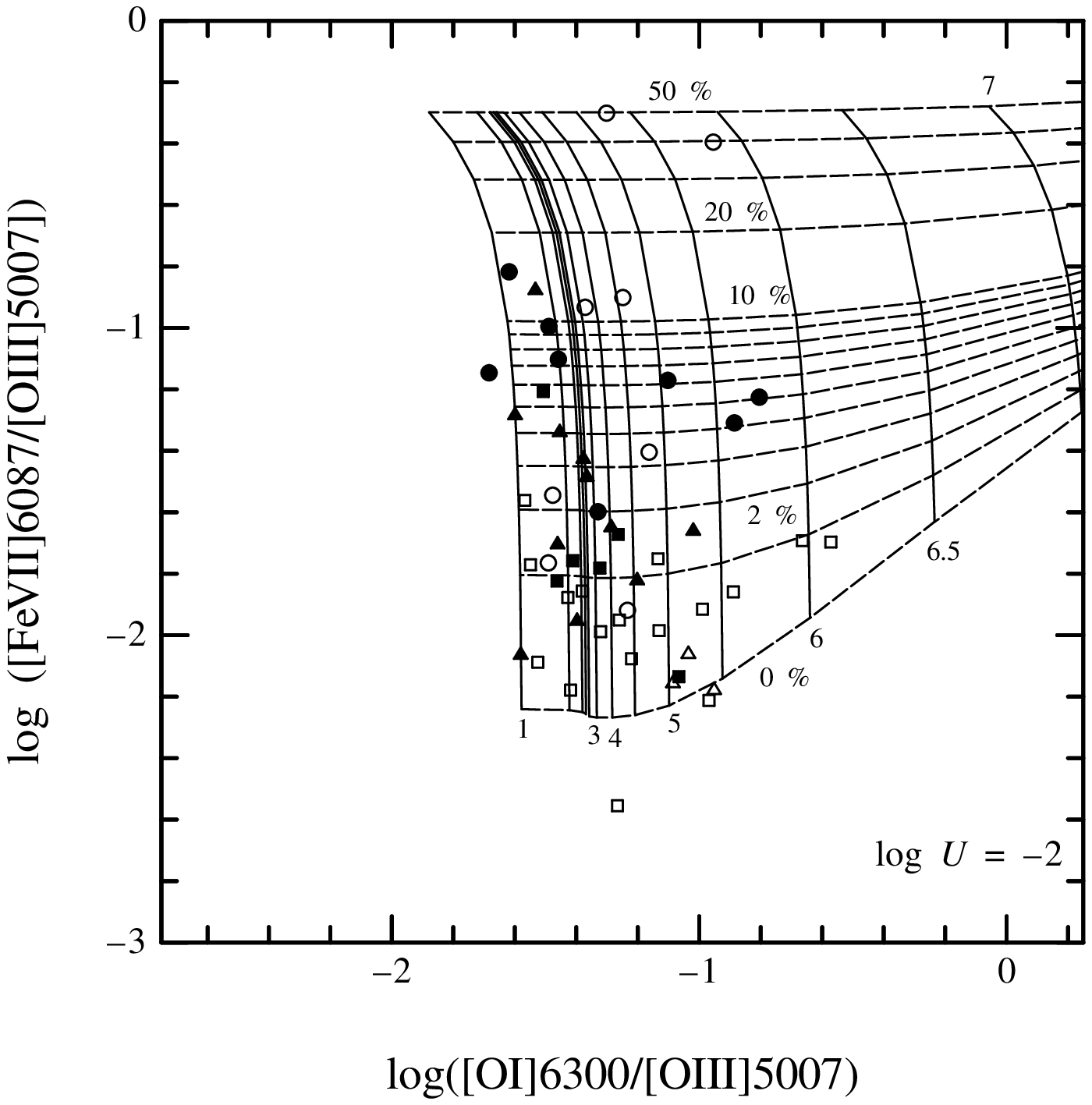}
\caption{Dual-component photoionization models are shown in the
diagram of [\ion{Fe}{7}]/[\ion{O}{3}] vs.\ [\ion{O}{1}]/[\ion{O}{3}].
The circles are S1s, the triangles are S1.5s, and the squares are S2s.
The filled symbols denote the objects with [\ion{Fe}{10}] emission,
while the open symbols denote the objects without [\ion{Fe}{10}].
The numbers labeling the lowest dashed line represent $\log n_{\rm H}$.
The percentages represent contribution of the HINER [\ion{O}{3}] flux
to the total [\ion{O}{3}] flux.
}
\end{figure*}

\section{
 New Mid-Infrared Diagnostic of the Dusty Torus Model
 for Seyfert Nuclei
}

\subsection{The New MIR Diagnostic}
The current unified model of active galactic nuclei (AGNs) 
has introduced the dusty torus
around the central engine (Antonucci 1993).
Therefore, it is urgent to study the basic properties of 
dusty tori (e.g., Pier \& Krolik 1992).
Utilizing the anisotropic property of dusty torus emission,
we propose a new MIR diagnostic
to estimate a critical viewing angle of the dusty torus
between type 1 and 2 AGNs.

Because of the anisotropic properties of
the dusty torus emission,
the emission at $\lambda <$ 10 $\mu$m is systematically stronger
in type 1 AGNs than in type 2s while that at $\lambda >$ 20 $\mu$m
is not significantly different between type 1 and type 2 AGNs.
Therefore the luminosity ratio between 3.5 $\mu$m and 25 $\mu$m is 
expected to be highly useful to distinguish between type 1 and 2 AGNs
(Figure 6).
Here we define the above ratio as
\[
R = \log 
\nu_{\rm 3.5 \mu m}~f_{\rm \nu_{3.5 \mu m}}/\nu_{\rm 25 \mu m}~f_{\nu_{\rm 25 \mu m}}.
\]

\subsection{Results \& Discussion}

\begin{figure*}
\figurenum{6}
\epsscale{1.8}
\plotone{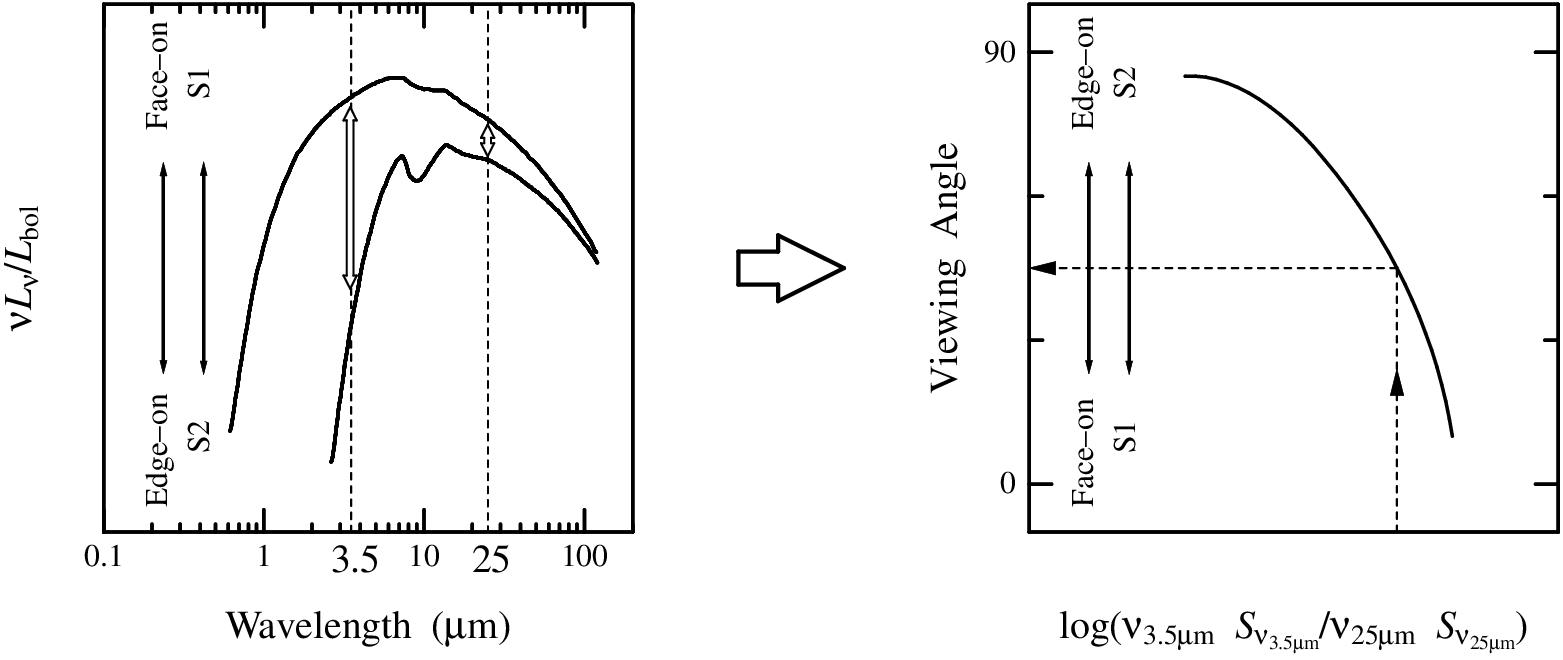}
\caption{
Basic concept of our MIR diagnostic. The upper panel shows
typical spectra of the torus emission for S1s (upper) and for S2s (lower).
The lower panel shows how the 3.5 \micron{} to 25 \micron{} flux ratio
yields the viewing angle toward the torus.
}
\end{figure*}

\begin{figure*}[p]
\figurenum{7}
\epsscale{1.0}
\plotone{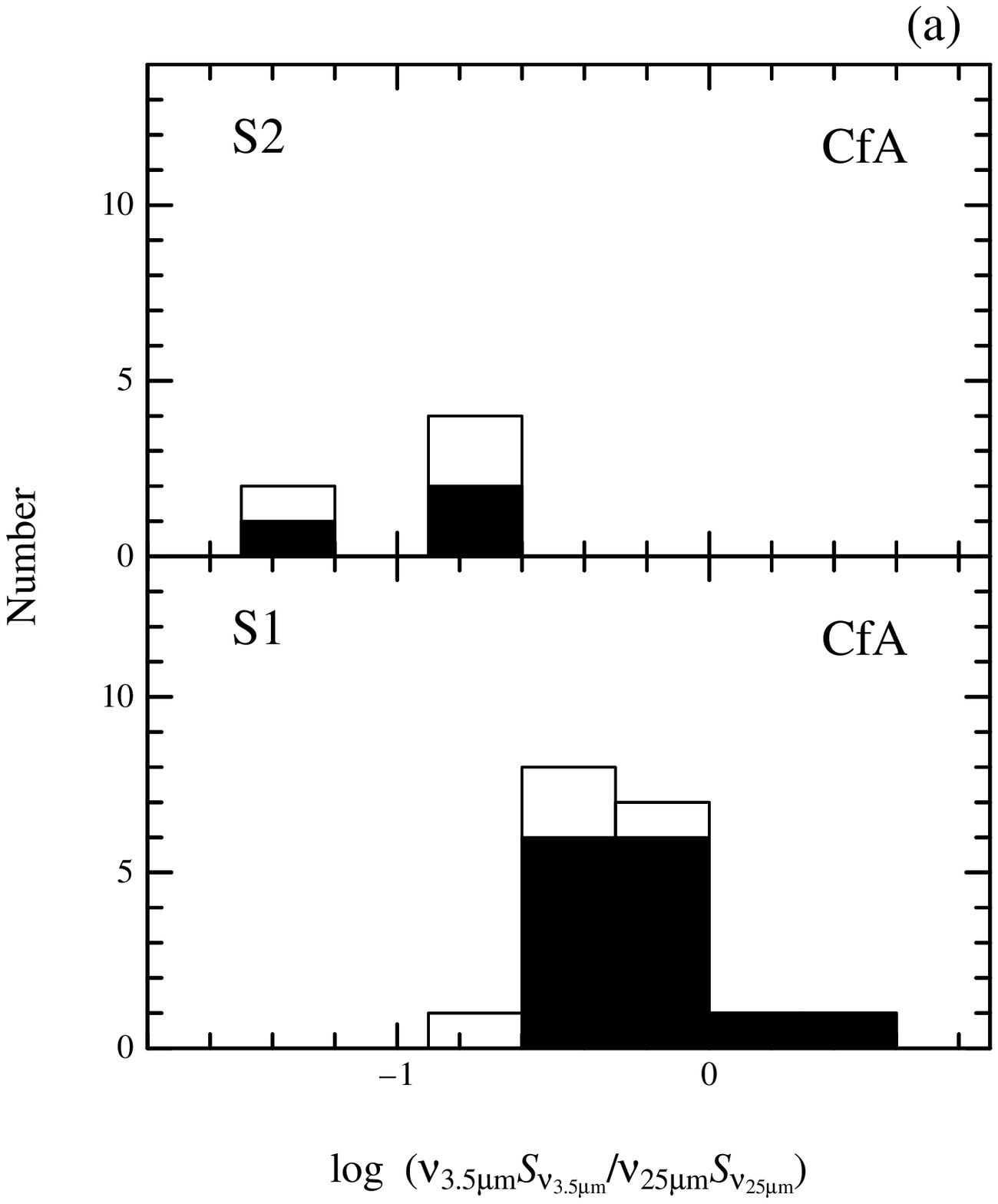}
\plotone{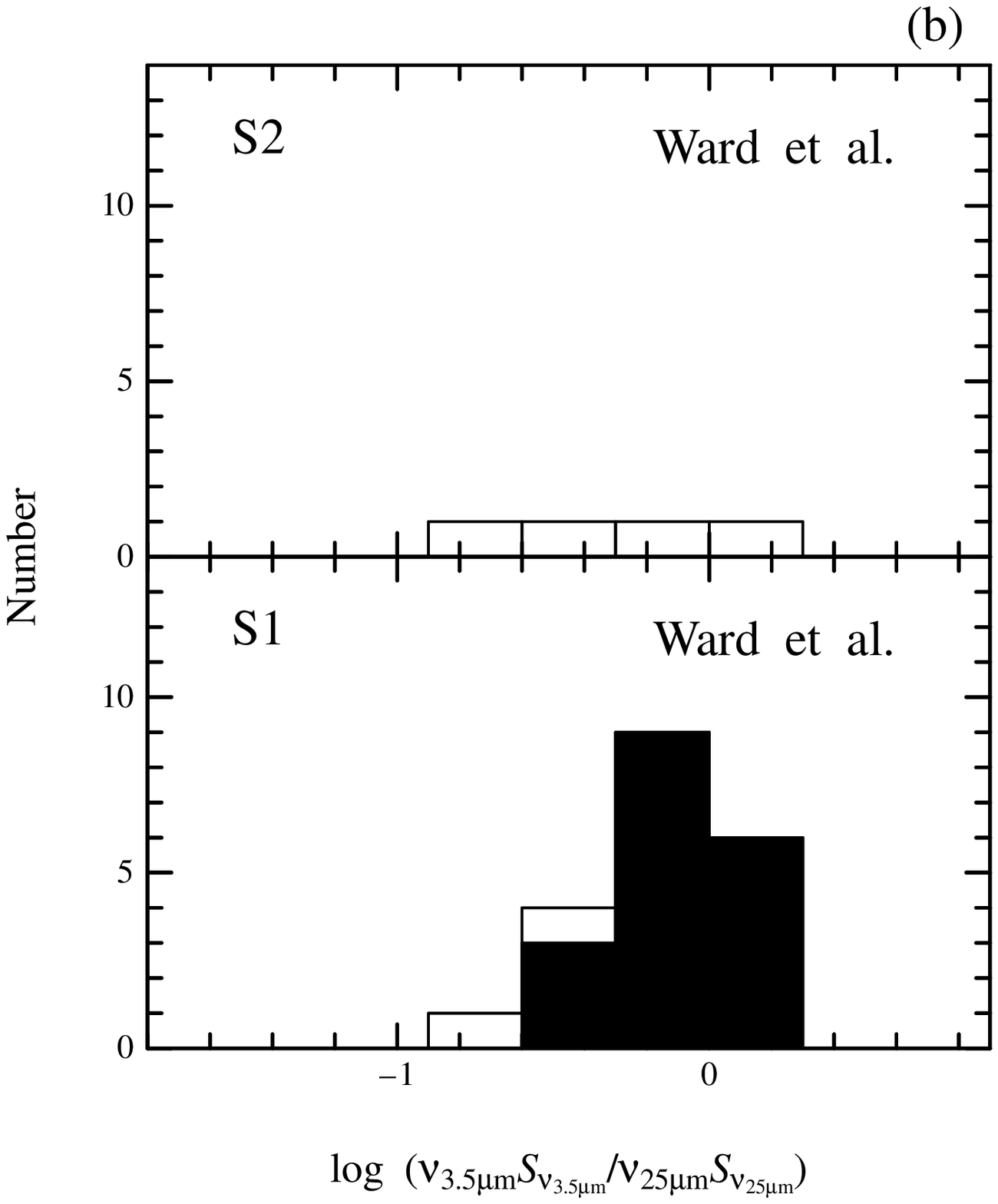}\\
\plotone{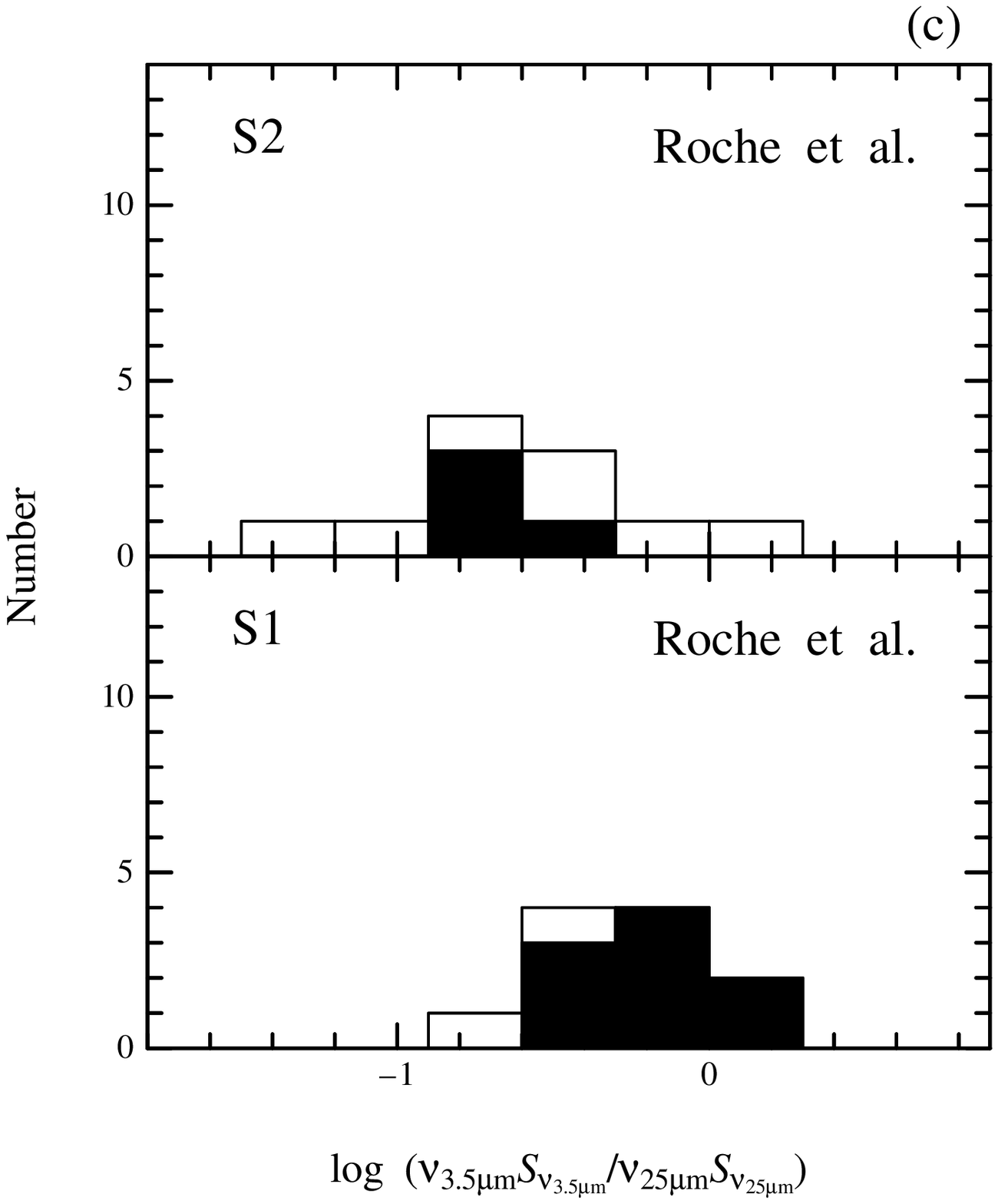}
\plotone{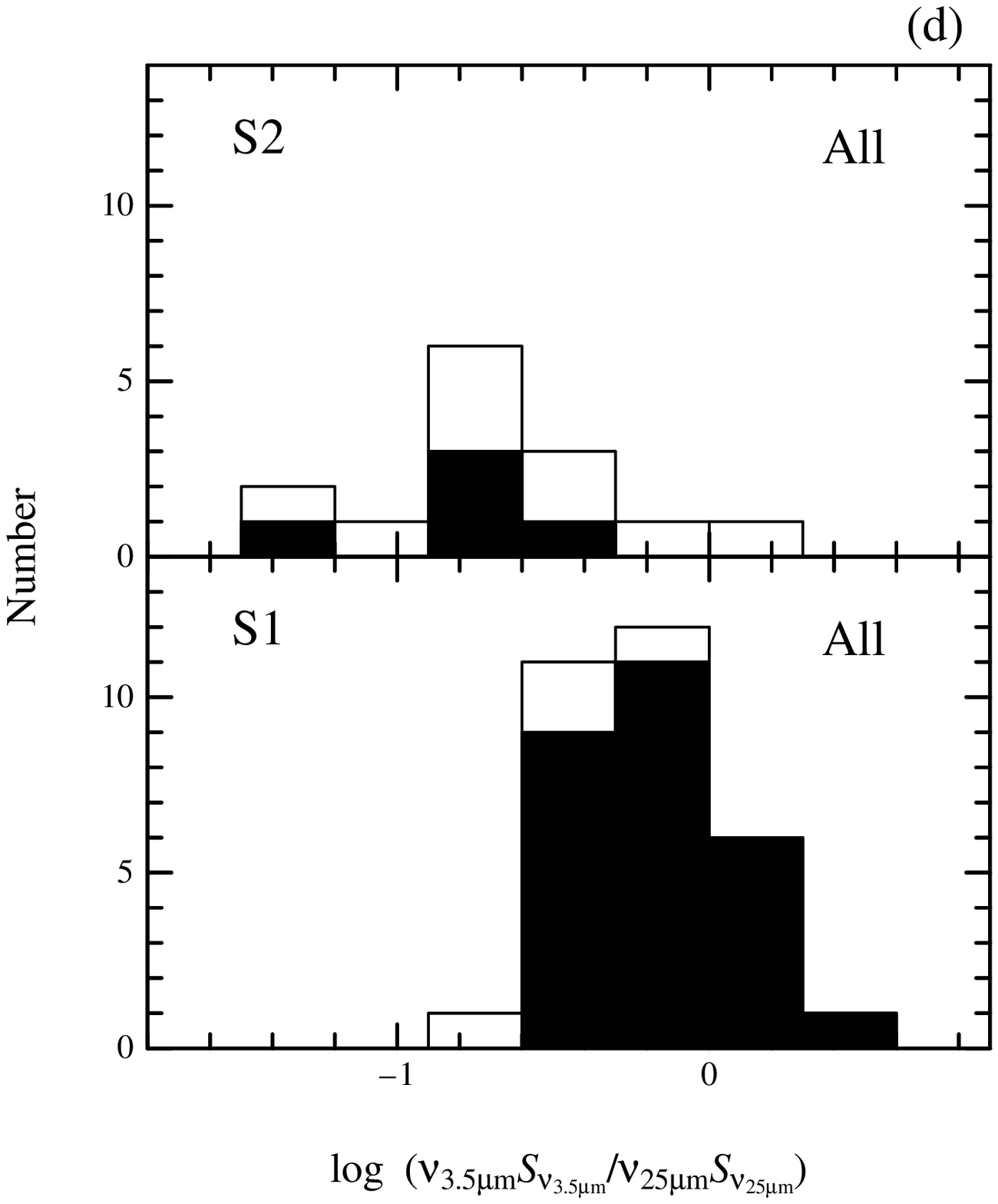}
\caption{
Histogram of $R(L,25)$ for the CfA Seyferts ($a$),
the sample of Ward et al.\ (1987) ($b$),
the sample of Roche et al.\ (1991) ($c$),
and the total sample ($d$). Galaxies shown by white bars are
likely to suffer from contamination and are not used in our analysis.
}
\end{figure*}

We  adopt three samples chosen by different selection criteria
and compiled photometric data in $L$, $N$, and {\it IRAS} 25 \micron{} bands:

\begin{enumerate}
\item 18 S1s and 6 S2s from the CfA Seyfert galaxies (Huchra \& Burg 1992)
\item 20 S1s and 4 S2s from the sample of Ward et al.\ (1987), which is
       limited by the hard X-ray flux from 2 to 10 keV 
\item 11 S1s and 11 S2s from the sample of Roche et al.\ (1991), which is
       composed of $N$-band bright objects
\end{enumerate}
Since some objects are included in more than one sample,
there are 31 S1s and 14 S2s in total.

The type 1 Seyferts are clearly distinguished from the type 2s
with a critical value $R\simeq -0.6$; $R > -0.6$ for type 1s while
$R < -0.6$ for type 2s (Figures 7a-d).
If we apply the Kolmogrov-Smirnov (KS) test, the probability that
the observed distributions of S1s and S2s originate in the same
underlying population
turns out to be 0.275 \%.

The upper panel of Figure 8 shows the theoretical models
of Pier \& Krolik (1992, 1993), which  are characterized
by $a$ (the inner radius of the torus), $h$ (the full height
of the torus), $\tau_{\rm r}$ (the radial Thomson optical depth),
$\tau_{\rm z}$ (the vertical Thomson optical depth),
and $T$ (the effective temperature of the torus) [see Figure 9].
The intersection of each model locus with $R=-0.6$ gives
a critical viewing angle. 
The critical viewing angle is expected to be
nearly the same as the typical semi-opening angle of the
ionization cones observed in Seyfert nuclei, $\simeq$ 30\arcdeg -- 40\arcdeg
(cf. Lawrence 1991 and references therein).
Figure 9 shows
that only two models give reasonable critical viewing angles,
$\simeq 46$\arcdeg -- 50\arcdeg though these values are slightly larger than
the semi-opening angle of the cone.
The model with $a/h$ = 0.3 may be suitable for tori in Seyfert nuclei
because this inner aspect ratio gives a semi-opening angle of the torus,
 $\simeq 30$\arcdeg, being consistent with those of the observed ionized cones.
Although there are some
contaminations from the host galaxies, circumnuclear starbursts, and
dust emission in the narrow-line regions,
the new diagnostic provides a powerful tool to
study the critical viewing angle.

\begin{figure*}
\figurenum{8}
\plotone{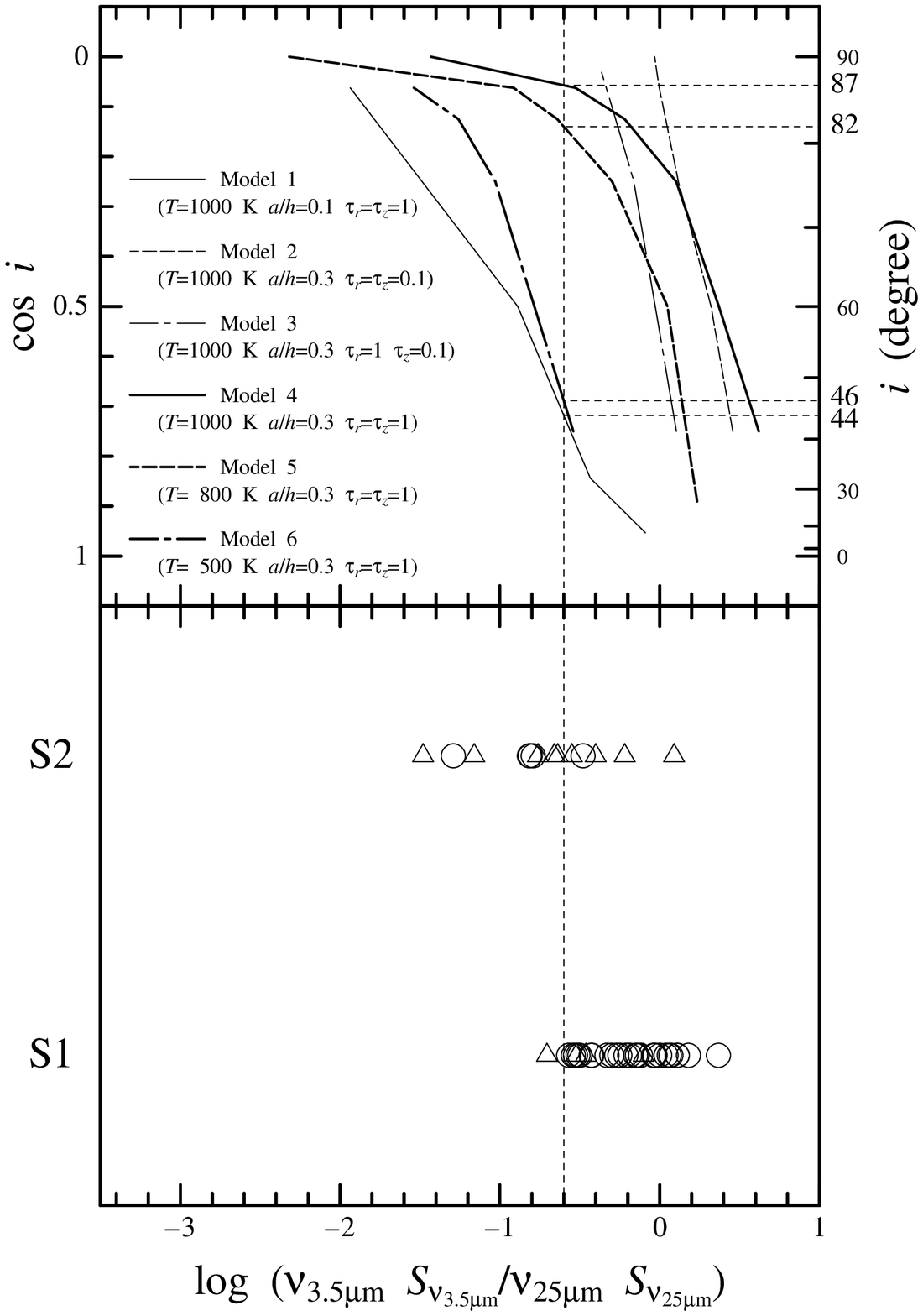}
\caption{
Upper panel: relationships between $R(L,25)$ and the viewing
angle for six dusty torus models given in Table 4. Lower panel:
distributions of the observed $R(L,25)$ values.
Galaxies shown by open triangles are likely to
suffer from contamination and are not used in our
analysis.
}
\end{figure*}

\begin{figure*}
\figurenum{9}
\plotone{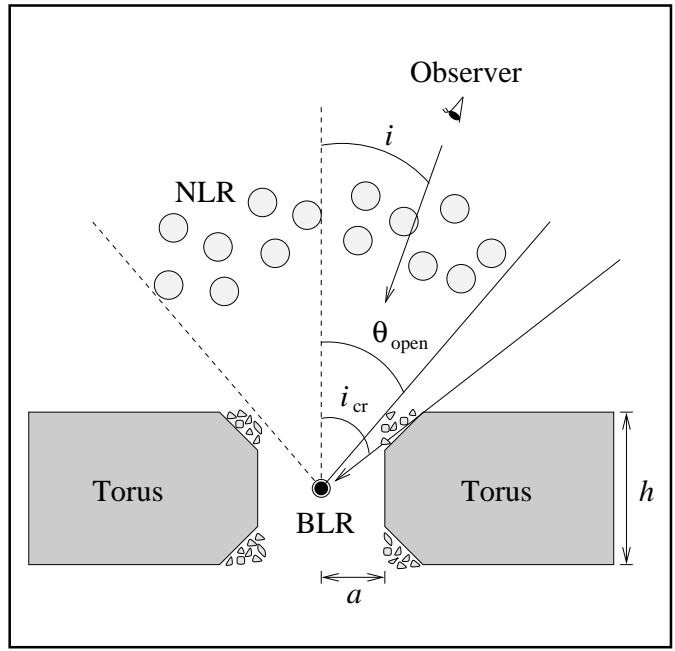}
\caption{
Schematic illustration of the geometrical configuration of the
dusty torus model.
}
\end{figure*}

\end{document}